\documentstyle[12pt]{article}

\date{}

\begin{document}

\title{{\LARGE\sf Spatial Inhomogeneity and Thermodynamic Chaos}}
\author{
{\bf C. M. Newman}\thanks{Partially supported by the
National Science Foundation under grant DMS-95-00868.}\\
{\small \tt newman\,@\,cims.nyu.edu}\\
{\small \sl Courant Institute of Mathematical Sciences}\\
{\small \sl New York University}\\
{\small \sl New York, NY 10012, USA}
\and
{\bf D. L. Stein}\thanks{Partially supported by the
U.S.~Department of Energy under grant DE-FG03-93ER25155.}\\
{\small \tt dls\,@\,physics.arizona.edu}\\
{\small \sl Dept.\ of Physics}\\
{\small \sl University of Arizona}\\
{\small \sl Tucson, AZ 85721, USA}
}
\maketitle

{\bf Abstract:}  We present a coherent approach to the competition between
thermodynamic states in spatially inhomogeneous systems,
such as the Edwards-Anderson spin glass with a fixed
coupling realization.  This approach explains
and relates chaotic size dependence, ``dispersal
of the metastate'', and for replicas:  non-independence,
symmetry breaking, and overlap (non-)self-averaging.

\bigskip
\bigskip

The connection between the existence of many thermodynamic states and
the phenomena of non-self-averaging (NSA) and replica symmetry
breaking (RSB) has long been a central topic of research
in disordered systems. These phenomena play a key role in
Parisi's solution \cite{Parisi1,Parisi2,Mezard} of the infinite-ranged
Sherrington-Kirkpatrick (SK) Ising spin glass model \cite{SK},
and have been discussed in many other contexts,
e.g., short-ranged spin glasses
\cite{Parisi3,FPV}, random field XY models \cite{LD}, random manifolds
\cite{MP},
heteropolymers \cite{BMY1}, and impure superconductors in magnetic
fields \cite{BMY2}. Another aspect of the competition between thermodynamic
states, introduced by the authors in Ref.~\cite{NS92}, is ``chaotic size
dependence'' (CSD) which, unlike NSA, manifests itself for a
fixed realization of the disorder.

An important issue is whether (and in what sense) the many novel
features of Parisi's solution can apply to more
realistic models, such as the Edwards-Anderson (EA)
nearest-neighbor Ising spin glass \cite{EA}.
In a previous paper
\cite{NS95}, it was shown that this ``SK picture'', as conventionally
understood,
is not valid.  In particular, the overlaps of the thermodynamic
states for coupling realization ${\cal J}$ do {\it not\/} depend
on ${\cal J}$.
This leads us to approach all basic phenomena as
accessible for a {\it fixed\/} realization.
Although we focus here on disordered systems, this approach is applicable
to the more general setting of inhomogeneous systems.

We shall see that CSD is one aspect of a
phenomenon, ``dispersal of the metastate'', which is closely connected
with both NSA and RSB. The metastate
is a natural ensemble, i.e., a probability measure, on
the (pure or mixed) thermodynamic states of the system \cite{AW}.
At high temperature, it
is a $\delta$-function on a single state; we call this a non-dispersed
metastate.  CSD will occur if the metastate ceases to be a single
$\delta$-function; we call this a dispersed metastate. It will
become apparent below, but we emphasize now, that {\it dispersal of
the metastate is not the same as the mere existence of many states\/}.
One unfamiliar but essential feature of our approach is a
replacement for the usual notion of NSA;
we shall see that (unlike the standard SK picture)
{\it replica overlap fluctuations (if they occur)
are due not to explicit ${\cal J}$-dependence
but rather to state-dependence within the metastate
for fixed ${\cal J}$}.

We also find further connections between the
structure of the metastate and that of the replicas: dispersal
of the metastate is equivalent to {\it replica non-independence\/}
(RNI), to be explained below, while RSB is equivalent to
the individual states in the metastate being mixtures (of pure
states).

\medskip

{\it CSD and the metastate.\/} --- A thermodynamic state $\Gamma$ of
a system such as the EA model at temperature T with fixed ${\cal J}$
(we take zero magnetic field)
is in general a mixture of pure (extremal) states. If there are many pure
states, e.g., as predicted by the SK picture \cite{MPV},
CSD could occur \cite{NS92}. This would mean that for
finite volume Gibbs states with ${\cal J}$-independent boundary
conditions, such as $\rho^{(L)}$ (with periodic b.c.'s on the $L^d$
cube $\Lambda_L$ centered at the origin), the correlations
$\langle \sigma_{i_1}\dots\sigma_{i_m}\rangle_L$ will have no single limit
as $L\to\infty$ but rather many different limits along different subsequences
of $L$'s.

Such behavior in $L$ is analogous to chaotic behavior in time $t$ along the
orbit of a dynamical system. In both cases, the dependence of the state
(on $L$ or $t$) is actually deterministic but appears to be a random
sampling from some distribution $\kappa$ on the space of states.
For inhomogeneous or disordered systems, $\kappa$ is a metastate,
i.e., a probability measure
on the space of all (fixed ${\cal J}$) thermodynamic states,
and is the limit of $\kappa_L$, a ``microcanonical'' ensemble in which each of
the finite
volume states $\rho^{(1)}, \ldots ,\rho^{(L)}$ has weight $L^{-1}$. The
meaning of the limit is that for every (nice) function $g$ on
states (i.e., a function of finitely many correlations),
\begin{equation}
\label{eq:macro}
\lim_{L\to\infty}L^{-1}\sum_{{\ell}=1}^{L} g(\rho^{({\ell})}) =
\{g(\Gamma)\}_\kappa\quad .
\end{equation}
Here, the bracket $\{ . \}_\kappa$ denotes the average over $\kappa$.

What are some possibilities for the metastate $\kappa$ in the
EA model? If there is a unique
thermodynamic state there is
no CSD and $\kappa$ is not dispersed -- it is a $\delta$-function
on that state. If
(as in the Fisher-Huse (FH) droplet picture of the spin glass phase
\cite{FH1,HF})
there is only a pair of pure states, related by a global spin flip, then
$\kappa$ is a $\delta$-function on the equal-weight mixture of these two
states; there is still no CSD and no dispersal of $\kappa$,
even though the
pure states break spin flip symmetry. On the other hand, a {\it dispersed\/}
metastate would result if the periodic b.c.'s in Eq.~(\ref{eq:macro})
were replaced by, say, plus b.c.'s -- namely, the equal-weight sum of the
two $\delta$-functions on each of the two pure states. The CSD yielding this
metastate simply corresponds to the strong preference, for most large
$L$'s, of plus b.c.'s for one or the other (chaotically depending on L)
of the two pure states. The possibilities in the EA model if there are
many pure states will be discussed later; these include both dispersed and
non-dispersed metastates.

\medskip

{\it What is NSA?\/} --- For disordered systems, there is
an alternate construction of metastates, due to Aizenman
and Wehr \cite{AW}. The microcanonical ensemble
$\kappa_L$ is replaced by the ensemble of states on $\Lambda_L$ obtained
by varying the couplings outside $\Lambda_L$. The limit here means that for
every (nice) function $F$ of finitely many couplings and finitely many
correlations,
\begin{equation}
\label{eq:macro2}
\lim_{L'\to\infty}[F({\cal J},\rho^{(L')})]_{av} =
\left[\{F({\cal J},\Gamma)\}_{\kappa({\cal J})}\right]_{av} \quad .
\end{equation}
Here, $[ . ]_{av}$ denotes the average over the (quenched) disorder
distribution $\nu$. The limit
exists, for subsequences of $L'$'s, and the resulting $\kappa({\cal J})$ is
a fixed-${\cal J}$ metastate \cite{AW}.  It can be shown
(we will present a proof elsewhere) that for
subsequences of ${\ell}$'s and $L$'s, the limit in Eq.~(\ref{eq:macro})
exists and yields the {\it same\/} $\kappa({\cal J})$. In both limits,
the subsequences are ${\cal J}$-independent. We conjecture that actually
all subsequences yield the same limiting $\kappa({\cal J})$, which would
then be {\it the\/} periodic b.c. metastate. For a different b.c., such
as plus, there could be a different metastate, as noted above.

When $F$ is chosen so as to fix the value of all couplings inside
$\Lambda_L$ and then $L\to\infty$, the RHS of Eq.~(\ref{eq:macro2}) reduces
to that of Eq.~(\ref{eq:macro}).  Averaging over $\kappa$ for fixed
${\cal J}$ thus corresponds to averaging over the ``couplings at infinity''.
In realistic systems, thermodynamic state observables can depend on the
bulk couplings and/or on the couplings at infinity. Thus
we suggest that {\it there are two
distinct types of dependence:
(i) on ${\cal J}$, and (ii) on the state $\Gamma$ within the
metastate $\kappa$ for fixed ${\cal J}$\/}.
We shall see that replica overlaps cannot have the first type of dependence,
but can in principle have the second kind.  If overlap fluctuations
(over the couplings) don't vanish
as $L\to\infty$ \cite{PS}, this is a signal that the {\it second\/} kind
of dependence holds for infinite volume.

\medskip

{\it Replica Non-Independence\/} --- Suppose the functions $F({\cal J},\rho)$
in the LHS of Eq.~(\ref{eq:macro2}) are restricted to be linear
in $\rho$, i.e., of the form $\langle f({\cal J},\sigma)\rangle_{\rho}$. These
determine a limit
for the finite volume pair $({\cal J}^{(L)},\sigma^{(L)})$, where
${\cal J}^{(L)}$ is the restriction of ${\cal J}$ to $\Lambda_L$
and $\sigma^{(L)}$ is distributed by $\rho^{(L)}$. The limiting
joint distribution is of the form $\nu({\cal J}) \rho_{\cal J}(\sigma)$
with $\rho_{\cal J}$ a fixed-${\cal J}$ thermodynamic state \cite{NS95,GKN},
which is simply the mean state (or barycenter)
of $\kappa({\cal J})$ \cite{AW} -- i.e.,
\begin{equation}
\label{eq:mean}
\langle\sigma_{i_1}\dots\sigma_{i_m}\rangle_{\rho_{\cal J}} =
\left\{\langle\sigma_{i_1}\dots\sigma_{i_m}\rangle_{\Gamma}\right\}_{\kappa({\cal J})}
\quad .
\end{equation}

Guerra has pointed out \cite{Guerra} an important extension
of this construction of $\rho_{\cal J}$.
Take uncoupled (but identical) Hamiltonians for infinitely many
replicas in finite volume, $\sigma^{1(L)},\sigma^{2(L)},\dots$.
This replaces $\rho^{(L)}(\sigma^{(L)})$ by the
product measure $\rho^{(L)}(\sigma^{1(L)})\rho^{(L)}(\sigma^{2(L)})\dots$, and
leads to a limiting joint distribution for
$({\cal J}^{(L)},\sigma^{1(L)},\sigma^{2(L)},\dots)$ of the form
$\nu({\cal J})\rho_{\cal J}^{\infty}(\sigma^1,\sigma^2,\dots)$.
Guerra further noted that nonvanishing of overlap fluctuations
(as $L\to\infty$) should imply the noninterchangeability of taking
(a) replicas and (b) the thermodynamic limit: $\rho_{\cal J}^{\infty}$
would {\it not\/} equal the product
$\rho_{\cal J}(\sigma^1)\rho_{\cal J}(\sigma^2)\dots$,
but rather some mixture of products.
In this case, one would have a Gibbs state
$\rho_{\cal J}^{\infty}$ for the uncoupled replica
Hamiltonians which is not simply the
product of Gibbs states for the individual ones.
The $\sigma^i$'s in such a $\rho_{\cal J}^{\infty}$
would not be independent,
but would be coupled implicitly through ``boundary conditions at infinity''.
We call this non-independence among replicas RNI.

We raise two additional points here. The
first is that replicas and the ``replica state'' $\rho_{\cal J}^{\infty}$
naturally follow (like $\rho_{\cal J}$) from the construction of the
metastate $\kappa({\cal J})$. The second is that
{\it RNI is equivalent to dispersal of the metastate $\kappa$
since the product decomposition of $\rho_{\cal J}^{\infty}$ is as
a mixture over $\kappa$\/}:
\begin{equation}
\label{eq:decomp}
\rho_{\cal J}^{\infty}(\sigma^1,\sigma^2,\dots) =
\{\Gamma(\sigma^1)\Gamma(\sigma^2)\dots\}_{\kappa({\cal J})}\quad .
\end{equation}

Both points can be explained in terms of ``metacorrelations''. Like the
usual correlations, $\langle\sigma_{A}\rangle_{\Gamma}$ (where $\sigma_A$
denotes
$\sigma_{i_1}\dots\sigma_{i_m}$ for the set $A = \{i_1,\dots,i_m\}$),
these are generalized moments of arbitrary order $m$ and they
characterize $\kappa$:
\begin{equation}
\label{eq:corr}
\{g(\Gamma)\}_{\kappa} = \left\{\langle\sigma_{A_1}\rangle_{\Gamma}\dots
\langle\sigma_{A_m}\rangle_{\Gamma}\right\}_{\kappa}\quad .
\end{equation}
Restricting to $m=1$ such $g$'s in Eq.~(\ref{eq:macro}) (or the
corresponding $F$'s in Eq.~(\ref{eq:macro2})) yields $\rho_{\cal J}$
as explained above. Restricting to $m\le2$ such $g$'s or $F$'s yields
the two-replica measure $\rho_{\cal J}^2(\sigma^1,\sigma^2)$ (corresponding
to ``integrating out'' all the other replicas in $\rho_{\cal J}^{\infty}$),
etc. Thus for any $m$, the correlation
$\langle\sigma_{A_1}^1\dots\sigma_{A_m}^m\rangle$
evaluated in $\rho_{\cal J}^{\infty}$ equals the LHS of Eq.~(\ref{eq:corr})
evaluated in $\kappa({\cal J})$. This proves
Eq.~(\ref{eq:decomp}).

Eq.~(\ref{eq:decomp}) shows that RNI and
RSB are distinct phenomena and either one can occur without the
other.  The former corresponds to a dispersal of $\kappa$
over multiple $\Gamma$'s, the latter to an individual $\Gamma$
being a mixture of multiple pure states.

\medskip

{\it Overlaps.\/} --- A basic construct in Parisi's solution \cite{MPV}
of the SK model is the overlap distribution. For configurations $\sigma,
\sigma'$ on all
space, the overlap should be defined as
\begin{equation}
\label{eq:overlap}
Q(\sigma,\sigma') =
\lim_{L\to\infty}L^{-d}\sum_{x\in\Lambda_L}\sigma_{x}\sigma'_{x}\quad .
\end{equation}
Its distribution depends of course on how $\sigma$ and $\sigma'$ are
chosen. If they are independently chosen from {\it pure\/} states \cite{MPV},
labelled $\alpha$ and $\beta$, then $\sigma_x \sigma'_{x}$ in
Eq.~(\ref{eq:overlap}) may be replaced by
$\langle\sigma_x\rangle_{\alpha}\langle
\sigma'_{x}\rangle_{\beta}$,
and $Q = q^{\alpha\beta}$ with distribution $\delta(q-q^{\alpha\beta})$.
In the standard SK picture,
$\sigma$ and $\sigma'$ are chosen from the
product measure $\rho_{\cal J}(\sigma)\rho_{\cal J}(\sigma')$.
Ref.~\cite{NS95} proved that this
picture is not valid for realistic models because the resulting overlap
distribution does not depend on ${\cal J}$.

However, Guerra's work \cite{Guerra} and that of the previous section indicate
that
the standard SK picture should be replaced by one where $\sigma$ and $\sigma'$
(and other replicas) are taken from $\rho_{\cal J}^{\infty}$. Before we
discuss various features of this nonstandard SK picture,
we will now see that {\it the overlap
structure still does not depend on ${\cal J}$\/}.

The overall overlap structure, given here by the (joint) distribution
$P_{\cal J}^{\infty}$ for overlaps, $Q^{ij} = Q(\sigma^i,\sigma^j)$,
between {\it all\/} pairs of replicas, contains much information.
The simplest piece of information is the (marginal) distribution
of a single overlap, say $q^{12}$; by Eq.~(\ref{eq:decomp}), this
equals $\{P_{\Gamma}(q^{12})\}_{\kappa({\cal J})}$, where
$P_{\Gamma}(q)$ denotes the overlap distribution coming from
$\Gamma(\sigma)\Gamma(\sigma')$. More generally,
the distribution for $Q^{12},Q^{34},Q^{56},\dots$ is
\begin{equation}
\label{eq:decomp2}
P_{\cal J}^{*}(q^{12},q^{34},\dots) =
\{P_{\Gamma}(q^{12})P_{\Gamma}(q^{34})\dots\}_{\kappa({\cal J})}\quad .
\end{equation}
This shows that $P_{\cal J}^{\infty}$ can encode considerable information about
the ${\cal J}$-dependent metastate $\kappa$, at least if different
$\Gamma$'s from $\kappa$ yield distinct $P_{\Gamma}$'s. Nevertheless,
$P_{\cal J}^{\infty}$ (and hence $P_{\cal J}^{*}$)
does not depend on ${\cal J}$.  The proof is essentially the same as
that given in Ref.~\cite{NS95}.

\medskip

{\it Scenarios for the EA metastate.\/} --- We previously discussed
the two possibilities for the EA model that the (periodic b.c.)
metastate $\kappa$ is 1) a $\delta$-function on a single pure state
or 2) a $\delta$-function on a mixture of two pure states.
If there are many pure states, other scenarios are
possible.  In all of these it
is convenient to imagine that for large $L$, $\rho^{(L)}$ is
approximately a mixture, $\sum_{\alpha}W_{L}^{\alpha}\rho^{\alpha}$,
of pure thermodynamic states labelled by $\alpha$ (we suppress the
fixed-${\cal J}$ subscript in this section); the scenarios
differ in how the weights $W_{L}^{\alpha}$ depend on $\alpha$
and $L$. Note that (for periodic b.c.'s) the ordered (i.e.,
with broken spin flip symmetry) pure
states always come in pairs with $W_{L}^{-\alpha}=W_{L}^{\alpha}$.

One scenario is that
$W_{L}^{*}\equiv\max_{\alpha}(W_{L}^{\alpha}+W_{L}^{-\alpha})\to0$
as $L\to\infty$.
Then each state $\Gamma$ from $\kappa$ should have an integral (rather
than sum) decomposition, $\Gamma = \int
W_{\Gamma}(\alpha)\rho^{\alpha}d\alpha$.
There are then two possibilities, both of which exhibit
RSB: either 3) there is no CSD (or RNI) and
$\kappa$ is a $\delta$-function on a single such $\Gamma$ or 4) there
is CSD (and RNI) and $\kappa$ is a dispersed measure over many such
$\Gamma$'s (and $W_{\Gamma}(\alpha)$'s). Both possibilities would be
quite unlike either the SK or FH pictures; competition between
pure states would be so well matched that for most large $L$'s,
no group of a few states dominates.

On the other extreme is the scenario where $W_{L}^{*}\to1$
as $L\to\infty$ (we assume here that every $\rho^{\alpha}$ is
ordered). This could occur without CSD (or RNI), which leads
back to possibility 2, or else 5) there is CSD (and RNI) and
$\kappa$ is dispersed over many $\Gamma$'s, each of the form
$[\rho^{\alpha}+\rho^{-\alpha}]/2$. This latter possibility is
an intriguing revision of the FH picture; the competition between
pure states would be so mismatched that for most large $L$'s, a
single pair of states would dominate, but which pair would depend
chaotically on $L$. Possibility 5 actually does occur in
the ground state structure of a short-ranged, highly disordered
spin glass model in high dimensions (while possibility 2 applies
in low dimensions) \cite{NS94}. Indeed, for the EA model itself at
$T=0$ (with a continuous distribution, such as Gaussian, for the
individual couplings), only possibilities 2 or 5 can occur.
We remark that in the plus b.c. version of possibility 5,
there would be RNI but no RSB.

The nonstandard SK picture corresponds to an intermediate
scenario where $W_{L}^{*}$ tends neither to zero nor to one and
most of the weight is concentrated on a few pure states, with the choice
of states depending chaotically on $L$. In this possibility 6),
there is CSD (and RNI) and $\kappa$ is a dispersed measure over
many $\Gamma$'s, each with a sum decomposition,
$\Gamma = \sum_{\alpha}W_{\Gamma}^{\alpha}\rho^{\alpha}$, so
there is also (nontrivial) RSB.
Such a $\Gamma$ immediately yields the (fixed-$\Gamma$) overlap distribution
\begin{equation}
\label{eq:SK}
P_{\Gamma}(q) =
\sum_{\alpha,\beta}W_{\Gamma}^{\alpha}W_{\Gamma}^{\beta}\delta(q-q^{\alpha\beta})
\quad .
\end{equation}

The key objects of this picture are the $P_{\Gamma}$'s and their
average over $\kappa$, $P\equiv\{P_{\Gamma}\}_{\kappa}$.
As noted, dependence on ${\cal J}$ and averaging over $\nu$ in
the standard SK picture are replaced by dependence on ${\Gamma}$ and averaging
over
$\kappa$. The basic requirements of
this nonstandard SK picture are as follows: The fixed-$\Gamma$ distribution
$P_{\Gamma}(q)$
should be a countable sum of (many) $\delta$-functions.  This is a
prerequisite for (nontrivial) ultrametricity within $\Gamma$, which is
the second requirement. The third requirement
is that the averaged distribution $P(q)$ be continuous between two
$\delta$-functions at $\pm q_{EA}$. We remark that the first and third
requirements cannot both be valid unless $P_{\Gamma}$ really does
depend on $\Gamma$.

The nonstandard SK picture is the only way in which some familiar aspects of
mean-field behavior can survive in the EA model, but it is not clear to us
whether this picture is in fact valid
in some dimensions at some temperatures.
We do know that the overlap structure
has no ${\cal J}$-dependence in both the standard and nonstandard
SK pictures. This ruled out the standard picture
\cite{NS95}; we now pursue some implications for the nonstandard
picture.

In addition to translation-covariance, the metastate $\kappa({\cal J})$
is covariant with respect to changes $\Delta H$ of finitely
many couplings (see Eq.~(5.3) of Ref.~\cite{AW}).  Under such changes, pure
states
remain pure and the pure state overlaps $q^{\alpha\beta}$ do not
change at all. However, the weights $W_{\Gamma}^{\alpha}$, which appear
in Eq.~(\ref{eq:SK}), will in general change.
The overlap structure of Eq.~(\ref{eq:decomp2}) yields a measure
on the set of weights appearing in the $P_{\Gamma}(q)$'s for each
fixed set of possible $q^{\alpha\beta}$'s and by the lack of
${\cal J}$-dependence, each of these measures must be invariant
under the change in weights created by every such $\Delta H$.  It is
unclear whether the enormous number of constraints this imposes
can be satisfied.

\medskip

{\it Discussion and conclusions.\/} --- For fixed
${\cal J}$, a metastate $\kappa = \kappa({\cal J})$ is a probability
measure on the fixed-$T$ thermodynamic states $\Gamma$ of the system.
The $\Gamma$'s can {\it a priori\/} be pure or mixed.
For disordered systems, metastates were constructed by Aizenman and
Wehr \cite{AW} by means of the ensemble of ``couplings at infinity''.
The average $\{\Gamma\}_{\kappa({\cal J})}$ over this metastate is
the state $\rho_{\cal J}$ (independently constructed in \cite{GKN})
used in \cite{NS95} to
rule out the standard SK picture for the EA model.

In this paper, we
constructed the same metastate $\kappa$, but by means of the
ensemble of finite volume states $\rho^{(L)}$ for a {\it fixed\/}
${\cal J}$, so that our approach
can be applied to inhomogeneous systems in general.
Dispersal of $\kappa$, if it occurs, implies
chaotic size dependence of $\rho^{(L)}$. Furthermore, replica
symmetry breaking is equivalent to the appearance of
mixed states $\Gamma$ in $\kappa$, while dispersal
of $\kappa$ is equivalent to replica non-independence
(as introduced by Guerra \cite{Guerra}).

We then classified the principal behaviors which
can occur, using the EA model as our prototype.
Among the scenarios are
possibilities 3 -- 5, which are
quite unlike any of the standard pictures.

Another new scenario (possibility 6),
referred to as the ``nonstandard SK picture'',
maximizes the features of mean field behavior that {\it might\/}
survive in short-ranged models despite the elimination
of the standard SK picture in Ref.~\cite{NS95}.
In this picture, the thermodynamic state
$\rho_{\cal J}^{\infty}$ on all replicas for
{\it fixed\/} ${\cal J}$ can be written as an
integral, over the $\Gamma$'s of
$\kappa({\cal J})$, of independent
$\Gamma$-distributed replicas (see Eq.~(\ref{eq:decomp})).
Each $\Gamma$ would be a weighted sum,
$\sum_{\alpha}W_{\Gamma}^{\alpha}\rho_{\cal J}^{\alpha}$,
from a {\it countably infinite\/} family of
pure states, but the integral (given by $\kappa$) would
be over a {\it continuum\/} of choices of these families
(as well as over the weights for each family), all
for a fixed ${\cal J}$.
The measure $\kappa$ prescribes the probability of
obtaining both a particular family of pure states and their weights.
This probability corresponds to selecting the family
(and weights) according to $\rho^{(L)}$ with
L large and chosen at random (but with ${\cal J}$ fixed);
each different choice of $L$ will in general select
out a different family of pure states.

If such a picture holds,
some familiar mean-field behavior would be exhibited
in short-ranged models. In particular, one
would find dominance of
a few states in large finite volumes (but
with the family from which they're chosen
changing chaotically with volume).
A restricted (within $\Gamma$) ultrametricity of overlaps
is not ruled out, but the countable
nature of each $\Gamma$ places
strong constraints on the nature of
the metastate.

We emphasize that the nonstandard SK picture
differs from the usual one in
important respects.  In particular,
there is no dependence of
overlap distributions on ${\cal J}$, but
only on the state $\Gamma$ within $\kappa$.
Also, ultrametricity would not
hold in general among any three pure states \cite{NS95},
but would be valid only for states from the same $\Gamma$.

We have previously shown \cite{NS95} that short-ranged models
display non-mean-field behavior; i.e.,
the existence of a ${\cal J}$-dependent state
$\rho_{\cal J}$ with ${\cal J}$-independent overlaps.
If some of the more familiar mean-field properties
are nevertheless to hold
for the thermodynamic states, then
something like our nonstandard SK picture must be present.
But this picture is
very heavily constrained. The
probability distribution on weights of pure states
and their overlaps {\it cannot change with\/}
${\cal J}$; in particular the distribution on weights must be invariant with
respect to any changes of any finitely many couplings.
Whether this is reasonable can be judged by the reader;
whether it will survive must be determined
by future work.

\medskip

{\it Acknowledgments.\/}  The authors thank M.~Aizenman and
A.~van~Enter for useful conversations; they also gratefully acknowledge
an extremely stimulating correspondence with F.~Guerra.

\end{document}